\documentclass{nature}

\usepackage{amsmath}

\usepackage{placeins}

\usepackage{booktabs}
\AtBeginDocument{
\heavyrulewidth=.08em
\lightrulewidth=.05em
\cmidrulewidth=.03em
\belowrulesep=.65ex
\belowbottomsep=0pt
\aboverulesep=.4ex
\abovetopsep=0pt
\cmidrulesep=\doublerulesep
\cmidrulekern=.5em
\defaultaddspace=.5em
}
\usepackage{multirow}

\usepackage{graphicx}
\usepackage[tablename={Extended Data Table}]{caption}
\usepackage{url}

\makeatletter
\let\saved@includegraphics\includegraphics
\AtBeginDocument{\let\includegraphics\saved@includegraphics}
\renewenvironment*{figure}{\@float{figure}}{\end@float}
\makeatother

\makeatletter
\renewenvironment{table}{%
  \renewcommand* {\@floatboxreset}{%
    \reset@font\@setminipage}
  \scriptsize	\@float{table}%
}{%
  \end@float\normalsize
}
\makeatother

\title{The Ultrafast Einstein--De Haas Effect}

\author{C.~Dornes$^{1}\textsuperscript{*}$, Y.~Acremann$^{2}$, M.~Savoini$^{1}$, M.~Kubli$^{1}$, M.~J.~Neugebauer$^{1}$, E.~Abreu$^{1}$, L.~Huber$^{1}$, G.~Lantz$^{1}$, C.~A.~F.~Vaz$^{3}$, H.~Lemke$^{3}$, E.~M.~Bothschafter$^{3}$, M.~Porer$^{3}$, V.~Esposito$^{3}$, L.~Rettig$^{3,4}$, M.~Buzzi$^{3,5}$, A.~Alberca$^{3}$, Y.~W.~Windsor$^{3,4}$, P.~Beaud$^{3}$, U.~Staub$^{3}$, Diling~Zhu$^{6}$, Sanghoon~Song$^{6}$, J.~M.~Glownia$^{6}$, S.~L.~Johnson$^{1}\textsuperscript{*}$}

\begin{document}

\maketitle

\begin{affiliations}
\item Institute for Quantum Electronics, Physics Department, ETH Zurich, CH-8093 Zurich, Switzerland
\item Laboratory for Solid State Physics, Physics Department, ETH Zurich, CH-8093 Zurich, Switzerland
\item Swiss Light Source, Paul Scherrer Institute, CH-5232 Villigen PSI, Switzerland
\item Fritz Haber Institute of the Max Planck Society, Faradayweg 4-6, D-14195 Berlin, Germany
\item Max Planck Institute for the	Structure and Dynamics of Matter, D-22761 Hamburg, Germany
\item Linac Coherent Light Source (LCLS), SLAC National Accelerator Laboratory, Menlo Park, California 94025, USA
\end{affiliations}
\newpage

\begin{abstract}
The original observation of the Einstein-de Haas effect was a landmark experiment in the early history of modern physics that  illustrates the relationship between magnetism and angular momentum\cite{Richardson1908,Einstein1915}.  Today the effect is still discussed in elementary physics courses to demonstrate that the angular momentum associated with the aligned electron spins in a ferromagnet can be converted to mechanical angular momentum by reversing the direction of magnetisation using an external magnetic field.  In recent times, a related problem in magnetism concerns the time-scale over which this angular momentum transfer can occur.  It is known experimentally for several metallic ferromagnets that intense photoexcitation leads to a drop in the magnetisation on a time scale shorter than 100~fs, a phenomenon called ultrafast demagnetisation\cite{Beaurepaire1996,Carpene2008,Faehnle2018}.  The microscopic mechanism for this process has been hotly debated, with one key question still unanswered:  where does the angular momentum go on these femtosecond time scales?  Here we show using femtosecond time-resolved x-ray diffraction that a majority of the angular momentum lost from the spin system on the laser-induced demagnetisation of ferromagnetic iron is transferred to the lattice on sub-picosecond timescales, manifesting as a transverse strain wave that propagates from the surface into the bulk.
By fitting a simple model of the x-ray data to simulations and optical data, we
estimate that the angular momentum occurs on a time scale of 200~fs and corresponds to 80\%
of the angular momentum lost from the spin system.  Our results show that interaction with the lattice plays an essential role in the process of ultrafast demagnetisation in this system.
\end{abstract}
\newpage
Broadly speaking, proposed mechanisms for ultrafast demagnetisation fall into two categories: spin-flip scattering mechanisms and spin transport mechanisms. The first category explains the demagnetisation process as a sudden increase in scattering processes that ultimately result in a decrease of spin order. These scattering processes can include electron-electron, electron-phonon, electron-magnon and even direct spin-light interactions. On average, such scattering must necessarily involve a transfer of angular momentum from the electronic spins to some other subsystem(s). Candidates include the lattice, the electromagnetic field, and the orbital angular momentum of the electrons. Numerical estimates and experiments using circularly polarised light strongly suggest that the amount of angular momentum given to the electromagnetic field interaction is negligible\cite{Koopmans2003}, and experiments using femtosecond x-ray magnetic dichroism (XMCD) indicate that the angular momentum of both electronic spins and orbitals decrease in magnitude nearly simultaneously\cite{Stamm2007,Stamm2010,Boeglin2010}. The only remaining possibility for a spin-flip induced change in angular momentum therefore appears to be a transfer to the lattice via spin-orbit coupling, but this remains to be experimentally verified.

The second category of proposed mechanisms relies on the idea that laser excitation causes a rapid transport of majority spins away from the excited region. Transport is significantly less efficient for minority spins, leaving behind a region of reduced magnetisation density\cite{Battiato2010}.  This mechanism has recently been addressed experimentally. One experiment has observed indirect demagnetisation of ferromagnets by photoexciting a neighbouring metallic layer\cite{Eschenlohr2013,Khorsand2014}, while other groups claim to observe spin injection from a photoexcited ferromagnetic layer into a deeper-lying, unpumped layer of another ferromagnetic or nonmagnetic metal\cite{Turgut2013,Melnikov2011}. If we consider the demagnetisation process as a result of  spin transport alone, the angular momentum of the entire system remains in the spin subsystem. In this case, the observed demagnetisation is interpreted as a local effect, arising from the laser-induced spatial inhomogeneity. On the contrary, studies have found sizeable ultrafast demagnetisation of thin ferromagnetic films on insulating substrates\cite{Schellekens2013}. In a transport model, it is hard to explain an almost unchanged demagnetisation process when the large bandgap of the substrate blocks the transport of carriers away from the excited region. From this it appears that although spin transport can contribute to ultrafast demagnetisation, there must also be sizeable contributions from spin-flip scattering.

\begin{figure}
 \includegraphics[width=0.85\textwidth]{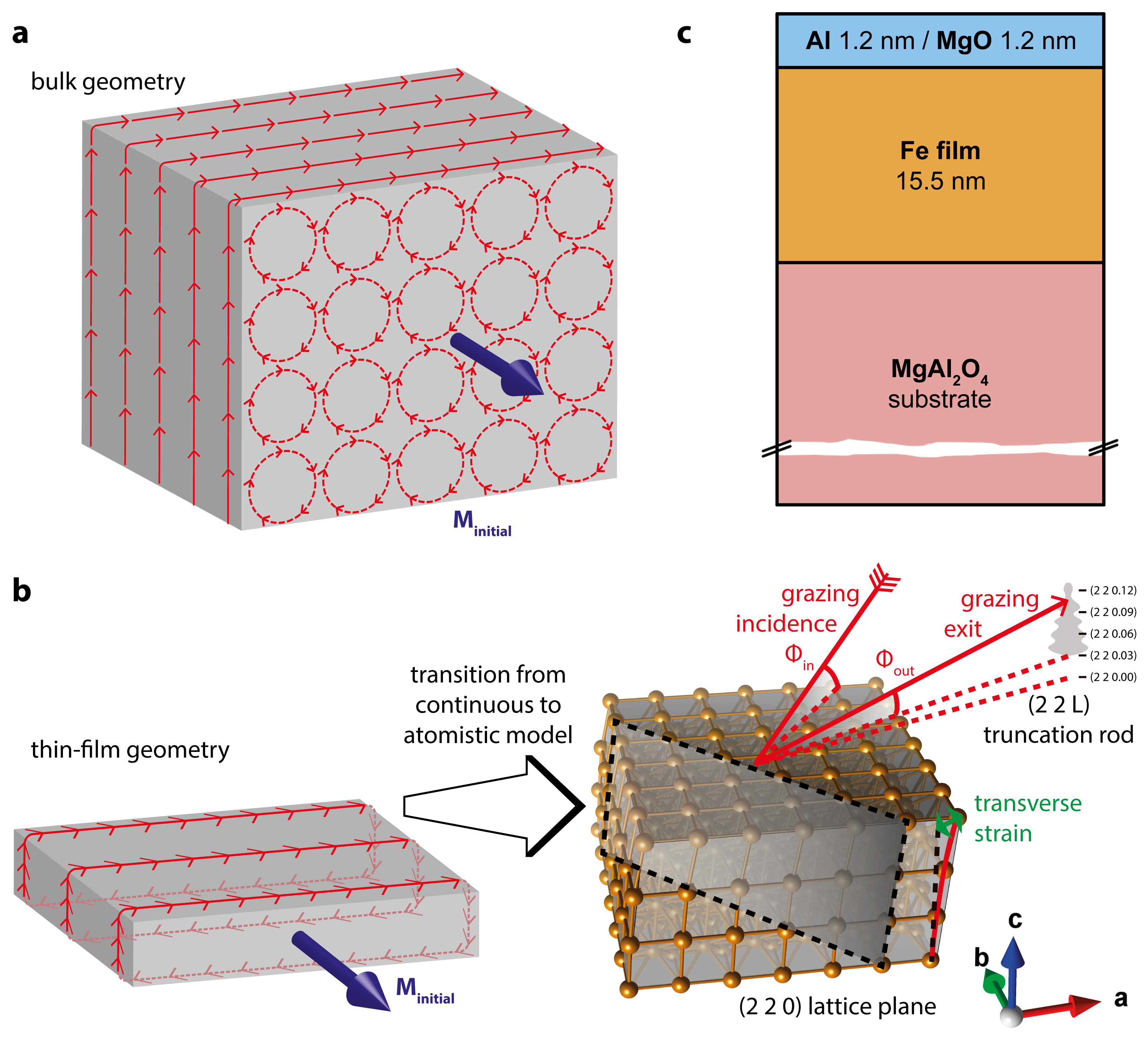}
 \caption{
\textbf{Overview of the idea of the experiment and its implementation.} The Gedankenexperiment of the ultrafast Einstein-De Haas effect is sketched in panel \textbf{a)}. In panel \textbf{b)}, on the left we show the corresponding situation in a thin-film sample, and on the right the actual geometry for our x-ray diffraction experiment. Panel \textbf{c)} contains the layer structure of the thin-film sample used in the experiment.
}\label{fig:gedanken}
\end{figure}

If indeed some fraction of angular momentum is transferred to the crystal lattice on ultrafast time scales, it should be possible to quantitatively measure the resulting structural dynamics.  The dynamics of the Einstein-de Haas effect have been previously considered in other contexts\cite{Jaafar2009,Mentink2018}.  Here we consider specifically the case where bulk, magnetised iron is uniformly demagnetised by a homogeneous femtosecond pump excitation.
The idea is sketched in Fig.~\ref{fig:gedanken}a.  The sudden demagnetisation causes mechanical torques which lead to unbalanced forces on the surfaces but not inside the bulk. More formally, we postulate that ultrafast demagnetisation causes a transient volume torque density $\vec{\tau} = - \gamma^{-1} d\vec{M}/dt$, where $\vec{M}$ is the initial magnetisation and $\gamma$ is the gyromagnetic ratio\cite{Einstein1915}. The latter is often expressed using the dimensionless magnetomechanical factor $g' = -\gamma \; \hbar / \mu_\mathrm{B}$, related but not identical to the spectroscopic $g$-factor\cite{Kittel1949,VanVleck1950}.  In a continuum model of lattice dynamics, a torque density $\vec{\tau}$ contributes to off-diagonal elements of an antisymmetric stress tensor $\sigma^M$ as $\sigma_{12}^M = -\sigma_{21}^M = \tau_3$, $\sigma_{23}^M = -\sigma_{32}^M = \tau_1$, $\sigma_{31}^M = -\sigma_{13}^M = \tau_2$. The resulting structural dynamics can be found by using the equation of motion
\begin{equation}
\rho {\partial^2  {u}_i \over \partial t^2} = \sum_{j} {\partial \sigma_{ij} \over \partial x_j},\label{eq:EoM_continuum}
\end{equation}
where $\rho$ is the mass density of the material, $u_i$ (with $i=1,2,3$) are  components of the displacement, $x_i$ are Cartesian coordinates and $\sigma_{ij}$ are components of the full stress tensor.  The stress tensor can be written as
\begin{equation}
\sigma_{ij} = \sum_{kl} C_{ijkl} \eta_{kl} + \sigma^M_{ij} + \sigma^\textrm{(ext)}_{ij}
\label{eq:stress}
\end{equation}
where $C_{ijkl}$ are the elastic constants, $\eta_{ij} = {\partial u_i \over \partial x_j}$ is the strain, and $\sigma^{(ext)}_{ij}$ are the components of any other applied stresses (e.g. expansive stress from heating, or magnetostriction) which do not constitute a torque density. Since Eq.~\ref{eq:EoM_continuum} contains only spatial derivatives of the stress, we expect the structural displacement caused by angular momentum transfer to begin where the gradient in the absolute demagnetisation is the largest, for example at interfaces.
As an illustrative example let us  consider the uniform demagnetisation of a free-standing thin iron film with surfaces normal to $z$. It is initially magnetised to saturation in the $x$-direction, as sketched in Fig.~\ref{fig:gedanken}b.  We assume that initially there is no strain in the film.  For now we also ignore effects such as thermal expansion, which would contribute to diagonal components of $\sigma^\textrm{(ext)}$.  Under these assumptions, the only non-zero  contributions in Eq.~\ref{eq:stress} are due to $\sigma^M_{23} = -\sigma^M_{32} = -\gamma^{-1} dM/dt$.  This implies that in the interior of the film the stress is uniform, with no immediate acceleration of the displacement.  At the free surfaces of the film we have the boundary conditions
\begin{equation}
\sigma_{3j} = 0,
\end{equation}
which allows us to simplify Eq.~\ref{eq:stress} to
\begin{equation}
\sum_{kl} C_{32kl} \, \eta_{kl} = {1 \over \gamma} {dM \over dt}.
\end{equation}
In $\alpha$-iron, the only non-zero components of the left-hand side involve $C_{3223}$, leading to
\begin{equation}
\eta_{23} = {1 \over \gamma}{1\over C_{3223}} {dM \over dt}
\end{equation}
at the two surfaces. Solving Eq.~\ref{eq:EoM_continuum} yields an impulsive transverse strain wave that propagates into the film from the interfaces.  The duration of strain wave generation is the timescale over which the angular momentum transfer occurs. For ultrafast timescales, the amplitude of the strain wave depends mainly on the total change of magnetisation---the information we seek---, as well as on the known density and elastic constants of the material. Expansive stresses from heating produce a similar effect via diagonal elements of $\sigma^\textrm{(ext)}$, but drive instead a longitudinal strain wave propagating from the interfaces, as is well established in ultrafast acoustics\cite{Thomsen1986}.

In our experiment we performed time-resolved x-ray diffraction measurements of these structural dynamics on a magnetic iron film. As illustrated in Fig.~\ref{fig:gedanken}c,  instead of a free-standing film we used a single-crystalline iron film grown epitaxially on a non-magnetic $\textrm{MgAl}_2\textrm{O}_4$ substrate with thin capping layers of MgO and Al.  The film is initially magnetised in-plane by an applied magnetic field, and is then partially demagnetised by an ultrafast laser pulse at 800~nm wavelength with 40~fs duration.  To measure the lattice dynamics with a high sensitivity to transverse strain, we perform grazing incidence diffraction close to the in-plane (2~2~0) reflection. Using an x-ray area detector, we map the intensity along its (2~2~L) crystal truncation rod as a function of time after excitation.  This intensity is directly related to the magnitude of the Fourier transform of the lattice displacements, and offers a direct measure of both transverse and longitudinal strain\cite{Macdonald1990}.

\begin{figure}
 \includegraphics[trim=0 0 -1.25cm 0cm,width=0.9\textwidth]{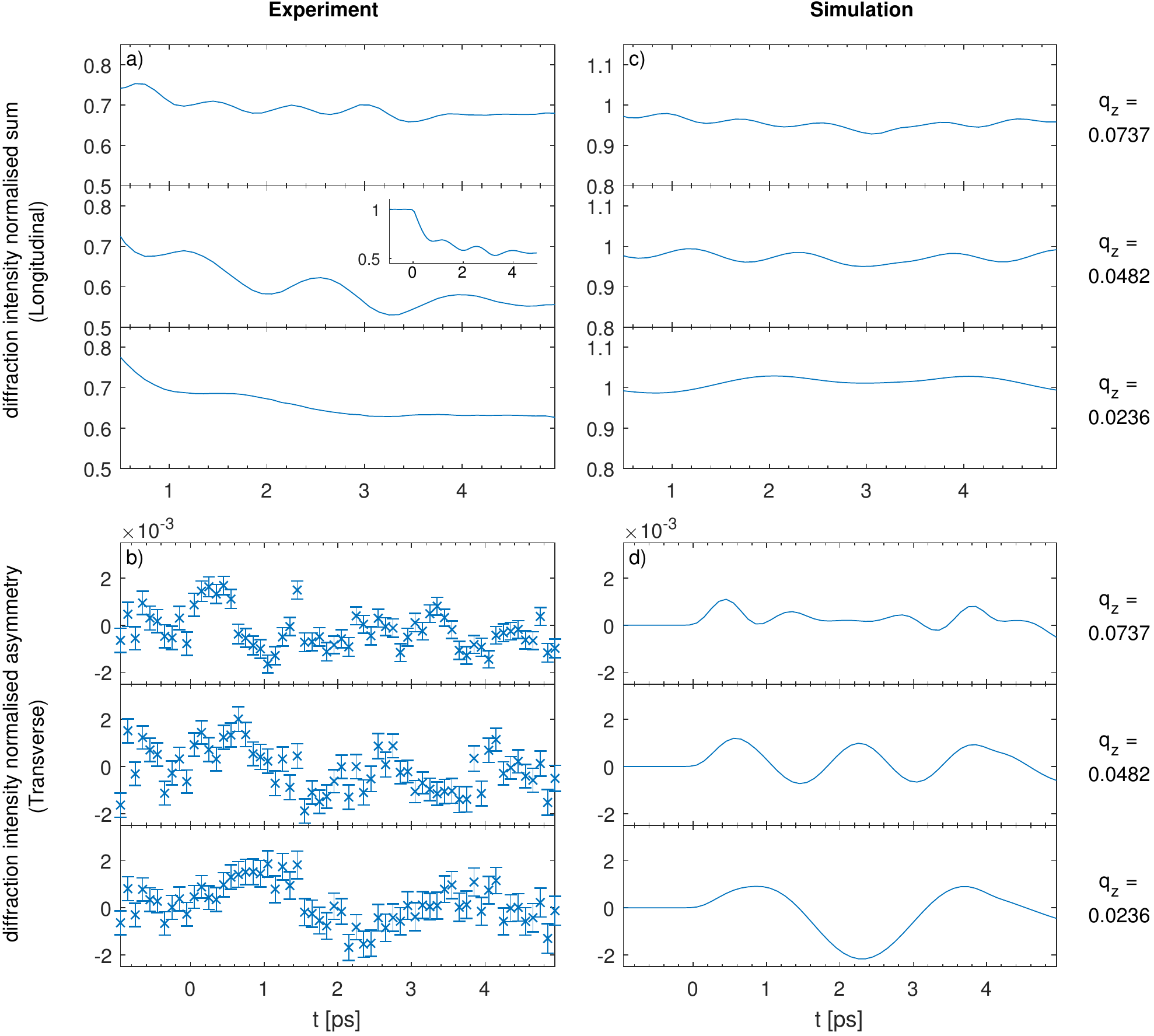}
 \caption{\textbf{Time-resolved x-ray diffraction signal, measured and simulated at different values of the out-of-plane momentum transfer $q_z$.} The central values of the three $q_z$-bands, in reciprocal lattice units, are labelled on the right side of the plot. Panels \textbf{a} \& \textbf{b} on the left show our experimental data; \textbf{c} \& \textbf{d} on the right show our best-fitting simulated data (absolute demagnetisation $\Delta M/M = 0.08$, demagnetisation time $200~\text{fs}$). The two upper panels \textbf{a} \& \textbf{c} show the sum of the intensities of both magnetisation directions, normalised to the static value at each $q_z$. For the experimental data in panel \textbf{a} the error bars were left out for clarity as they are on the order of the line thickness, and an inset of the full time and intensity range was added to the middle plot to illustrate the inititial intensity drop due to the Debye-Waller factor. The bottom two panels \textbf{b} \& \textbf{d} show the difference signal of the two intensities with the same normalisation per $q_z$.
}\label{fig:results}
\end{figure}

Fig.~\ref{fig:results} shows the data for selected values of the x-ray momentum transfer along the truncation rod alongside the results of simulations using a discrete model implementation of the lattice dynamics (see Methods section).  In order to identify dynamics that are connected to $\sigma^M$, we performed measurements for opposite directions of the initial magnetisation and subtracted the resulting data, a procedure which removes all contributions that do not depend on the sign of $d\vec{M}/dt$.  We see clear oscillations in this difference signal (Fig.~\ref{fig:results}b) which correspond closely to the expectations from our simulations.  This is direct evidence of angular momentum transfer to the lattice on sub-picosecond timescales.  For comparison, Fig.~\ref{fig:results}a also shows the dynamics for the sum of the data from opposite initial magnetisations.  These data show a drop in intensity (see inset in Fig.~\ref{fig:results}a, as well as oscillations at a slightly higher frequency.  The overall drop in intensity is likely to be an effect of a strong increase in thermal disorder\cite{Mohanlal1979} (Debye-Waller factor) which is not directly modelled in our simulations.  The oscillations have a period corresponding to that of longitudinal vibrational modes at the wavevector given by the momentum transfer $q_z$ along the rod, which is modelled in the simulations as the result of a longitudinal expansive stress that turns on with the laser excitation.

\begin{figure}
 \includegraphics[width=0.5 \columnwidth]{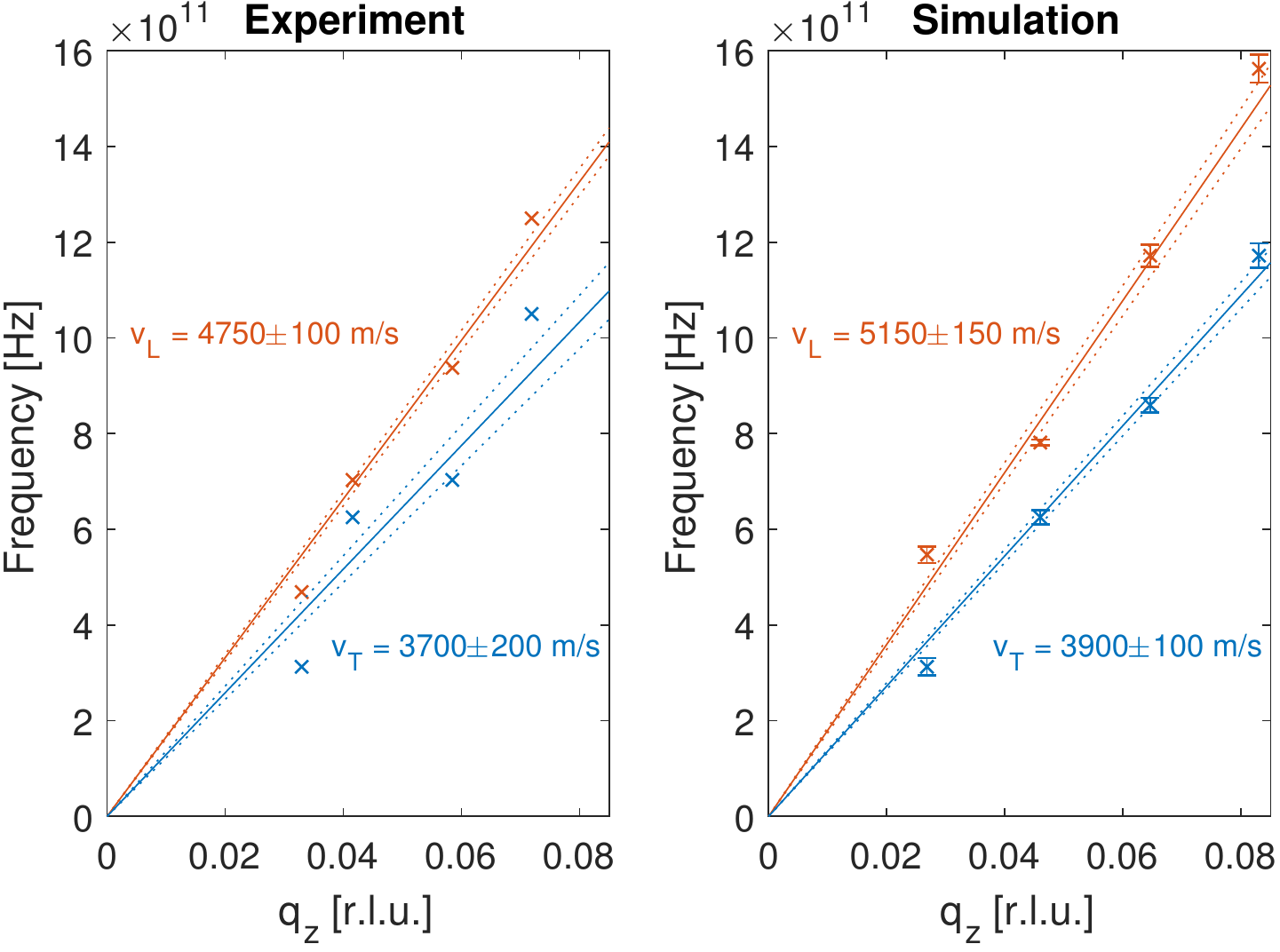}
 \caption{
\textbf{Frequency of modulations observed in the x-ray diffraction signal as a function of the out-of-plane momentum transfer $q_z$.}  The left panel shows the results from the experiment, while the right panel shows the results of an equivalent analysis on the simulated data sets.  The orange data points are derived from an FFT analysis of data resulting from summing the two opposite magnetisation directions, whereas the blue data points are derived from  subtracting data from the two opposite magnetisation directions (see Methods for details).  The solid lines show a fit to the general form $\omega = v q_q$, while the 1-$\sigma$ confidence interval of the fit is indicated with dashed lines.  Both experiment and simulation show a different slope $v$ for the fits to the sum and difference data with values consistent with the speed of sound for longitudinal and transverse acoustic waves, respectively.
}\label{fig:acoustic}
\end{figure}

Fig.~\ref{fig:acoustic} shows the frequency of oscillations as a function of wavevector $q_z$ for both the sum and difference data. This serves to confirm the presence of a transverse and longitudinal strain wave in an independent way that does not depend on our simulations. As expected, the difference data agrees with a linear relationship $\omega = v_T\,q$ where $ v_T = 3700 \pm 200~\text{m}/\text{s}$ is close to literature values of the transverse acoustic branch in [001]-direction\cite{Minkiewicz1967}, $3875 \pm 20~\text{m}/\text{s}$. The sum data instead show agreement with a linear relationship $\omega = v_L\,q$ with $v_L = 4750 \pm 100~\text{m}/\text{s}$, which is in agreement with our measurements of the longitudinal sound velocity by time-resolved optical reflectivity ($4730\pm150~\text{m}/\text{s}$, see Methods).

Based on fitting the simulations to the data, we can estimate both the timescale and magnitude of angular momentum transfer to the lattice. The simulations plotted in Fig.~\ref{fig:results} used the best parameters we found, which is a demagnetisation time of 200~fs and a total reduction in magnetisation of 8\%. On the same sample and under the same photoexcitation conditions, we observed a demagnetisation of 10\% using time-resolved magneto-optic Kerr spectroscopy. Setting the two values into relation, according to our best fit, the angular momentum transfer to the lattice due to the ultrafast Einstein--de Haas effect accounts for 80\% of the angular momentum loss seen optically.

Our results show conclusively that ultrafast demagnetisation involves a sub-picosecond transfer of angular momentum to the lattice. At the microscopic level, our results provide strong evidence for the importance of spin-flip processes that either directly or indirectly result in the emission of phonons.  We expect that similar dynamics govern the behaviour of other systems where an ultrafast change in spin angular momentum is observed, such as in Ni or Co, and in ferrimagnetic systems displaying all-optical switching\cite{Stanciu2007,Lambert2014,Takahashi2016,Granitzka2017}, where a rapid reversal of the direction of the magnetic moment is driven by femtosecond optical pulses. In the latter case, in addition to spin transport between magnetic sublattices or different regions of the material, our results indicate that lattice interactions may need to be considered for an adequate description of the ultrafast switching processes. A better understanding of the physical mechanisms behind this angular momentum transfer to the lattice can ultimately assist in the search for next-generation materials to use all-optical switching in practical devices.

\providecommand{\noopsort}[1]{}\providecommand{\singleletter}[1]{#1}%


\begin{addendum}
 \item Time-resolved x-ray diffraction measurements were carried out at the XPP endstation at LCLS. Use of the Linac Coherent Light Source (LCLS), SLAC National Accelerator Laboratory, is supported by the U.S. Department of Energy, Office of Science, Office of Basic Energy Sciences under Contract No. DE-AC02-76SF00515. Preparatory static diffraction measurements were performed at the X04SA beam line of the Swiss Light Source. We acknowledge financial support by the NCCR Molecular Ultrafast Science and Technology (NCCR MUST), a research instrument of the Swiss National Science Foundation (SNSF). E.A. acknowledges support from the ETH Zurich Postdoctoral Fellowship Program and from the Marie Curie Actions for People COFUND Program. E. M. B. acknowledges funding from the European Community's Seventh Framework Programme (FP7/2007-2013) under grant agreement n.290605 (PSI-FELLOW/COFUND). M. P. acknowledges supported by the NCCR MARVEL, funded by the SNSF.
 \item[Author contributions] C.D. and S.L.J. had the idea for and designed the experiment. C.A.F.V. performed the sample fabrication. C.D., M.S., M.K., M.J.N., E.A., L.H., E.M.B., M.P., V.E., L.R. and Y.W.W. performed synchrotron measurements to characterise the sample and prior sample candidates. M.B. and C.D. built the electromagnet. Y.A. built and programmed the pulser for the magnet. A.A. gave assistance to C.D. in analysing x-ray reflectivity measurements of sample candidates. M.S., C.A.F.V., L.H., E.A., G.L., H.L., M.B., P.B, and U.S. gave input to C.D. and S.L.J. on the experimental design and during data analysis. C.D., Y.A., M.S., M.K., H.L., E.M.B., M.P., U.S. and S.L.J. performed the experiment with the help of D.Z., S.S. and J.M.G., the LCLS beamline staff. C.D. and S.L.J. wrote the manuscript and all authors contributed to its final version.
 \item[Competing interests] The authors declare that they have no
competing financial interests.
 \item[Correspondence] Correspondence and requests for materials
should be addressed to C.D.~(email: dornesc@phys.ethz.ch) or S.L.J.~(email: johnson@phys.ethz.ch).
\end{addendum}

\FloatBarrier

\FloatBarrier

\begin{figure}
 \includegraphics[width=0.45\textwidth]{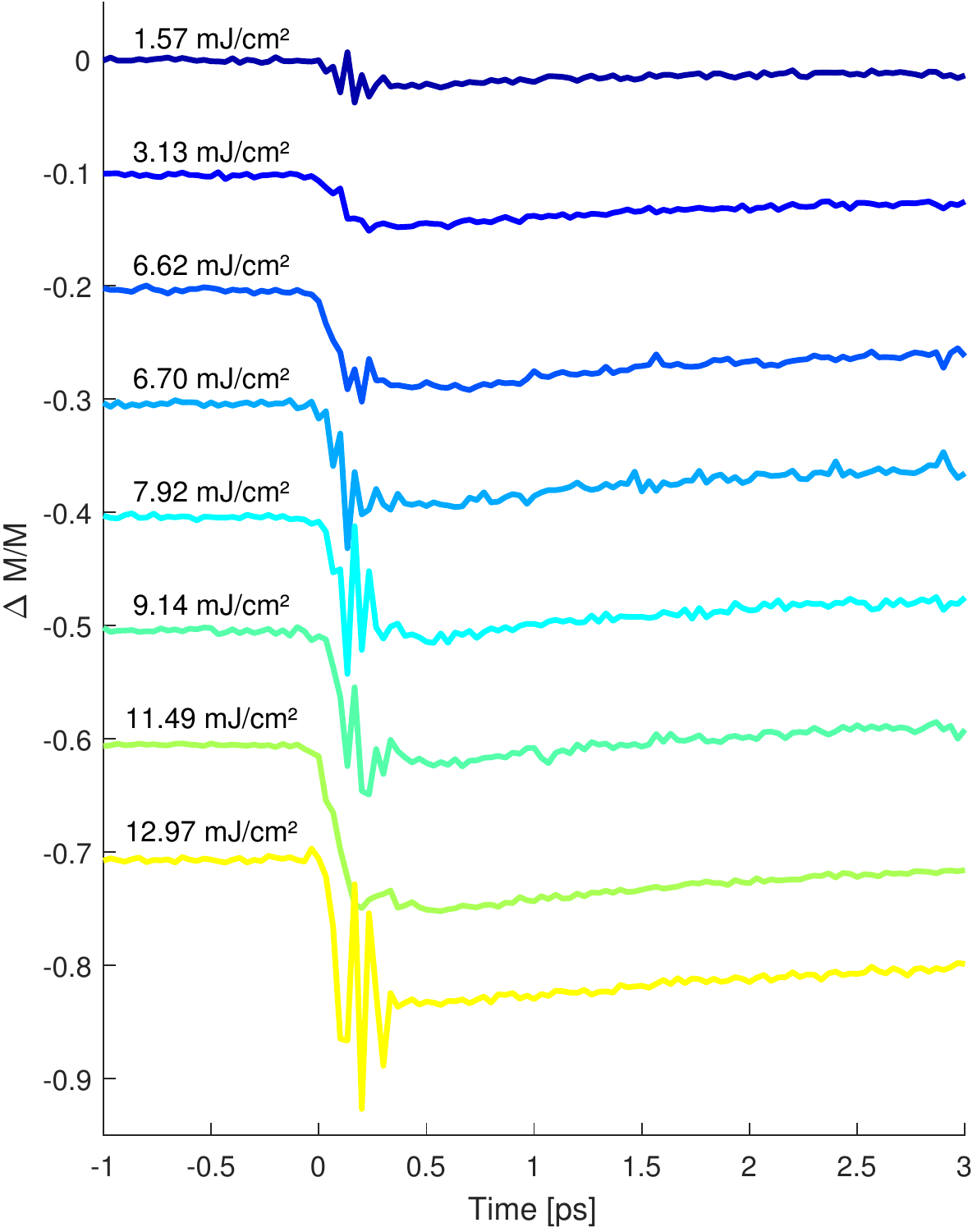}
 \caption{\textbf{Optical pump-probe measurements of the sample magnetisation at varying incident fluence.} The individual traces have been offset for clarity. These results were taken recreating the same geometric conditions as in the x-ray experiment, i.e. the laser spot size and angle of incidence were identical. At the highest fluence, laser damage was evident and also the static MOKE signal of the damaged area was permanently lower afterwards.}
 \label{fig:fluencetraces}
\end{figure}

\begin{figure}
 \includegraphics[width=0.55\textwidth]{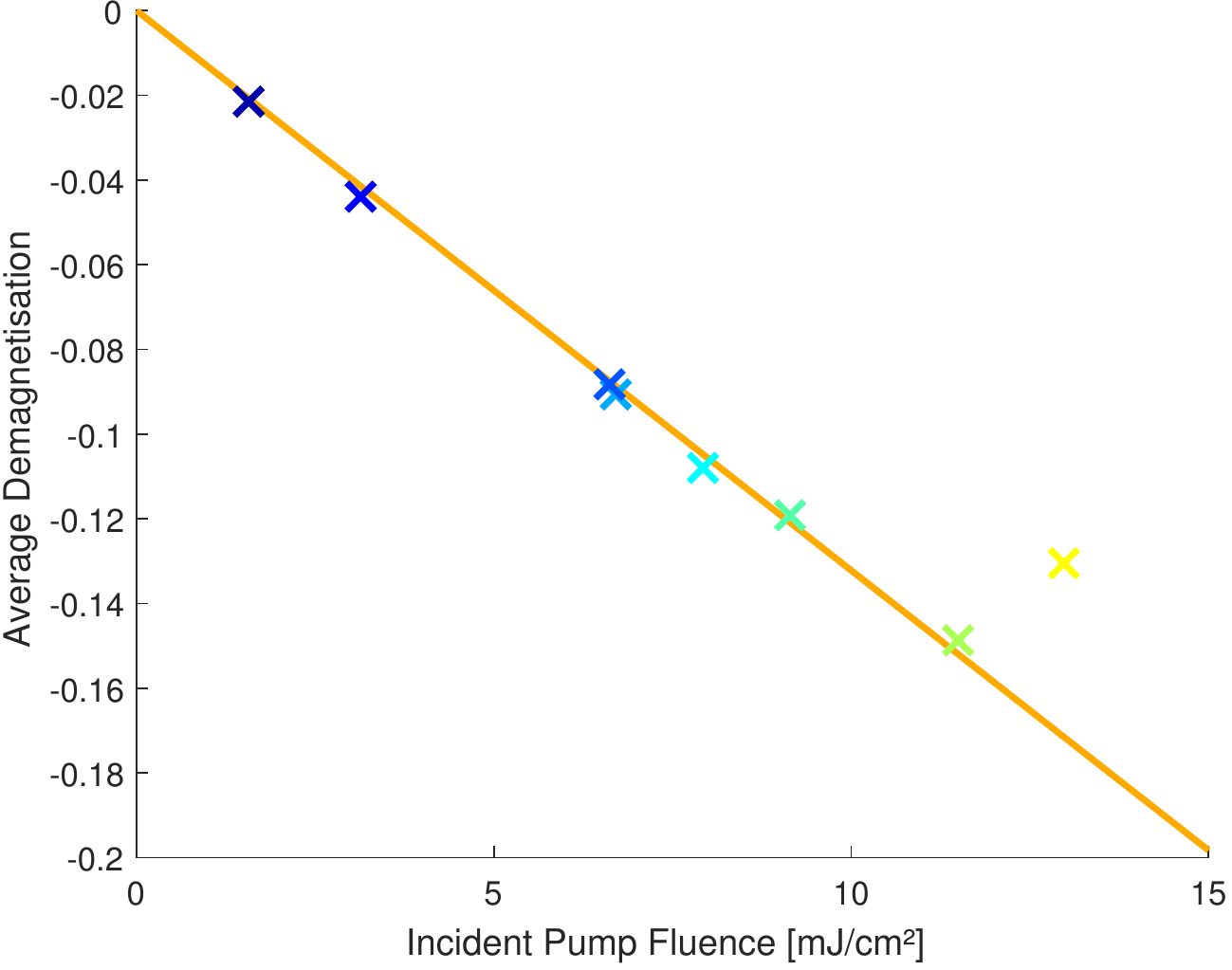}
 \caption{\textbf{Maximum relative demagnetisation as a function of fluence.} From the traces in Fig.~\ref{fig:fluencetraces}, the magnitudes of demagnetisation just after the coherent artifact (0.4~ps--0.7~ps) were extracted. A linear fit, excluding the highest fluence where damage was evident, captures the behaviour well. The fluence in the x-ray experiment was $8.0\pm0.3\text{mJ}/\text{cm}^2$, corresponding to a demagnetisation of 10\%.}
 \label{fig:fluencedependence}
\end{figure}

\begin{methods}

\section*{Sample preparation}

The single crystalline iron film with surface normal (0~0~1) was grown by molecular beam epitaxy on a $\text{MgAl}_2\text{O}_4$ (0~0~1) spinel substrate, capped on the top by nominally 2 nm layers of MgO and Al to prevent oxidation and x-ray damage. The $a$-$b$ in-plane axes of the film are rotated by $45^\circ$ with respect to the substrate ($\text{Fe}[110]{\parallel}\text{MgAl}_2\text{O}_4[100]$ epitaxial relationship). The high quality of the sample was verified by various methods, including in-situ low-energy electron diffraction (LEED), synchrotron measurements of the x-ray reflectivity (XRR) and the sample structure (unit cell parameters and lateral coherence length), as well as post-experiment TEM nanographs.  The TEM nanographs indicate a total thickness of 17.9~nm, whereas the XRR measurements show an Fe thickness of 15.5 nm with capping laters of 2.3~nm MgO and 2.7 nm Al.  Since the XRR fits have considererably less uncertainty on the Fe film thickness compared to the thin capping layers, we consider the discrepancy between TEM and XRR to indicate primarily an uncertainty on the thickness of the capping laters in the XRR measurements.

\section*{Transient Magneto-Optic Kerr Effect measurements}

To calibrate the magnitude of the demagnetisation of the particular sample used for the x-ray experiment, we performed transient magneto-optic Kerr effect (MOKE) measurements using another femtosecond laser operating at 800~nm. The beam profile and incidence angle of the pump from the x-ray experiment were reproduced precisely (see Methods section ``X-ray pump-probe experiment''); the pump repetition rate for the MOKE was 500~Hz and the pulse duration was 100~fs.  Figure~\ref{fig:fluencetraces} in the Extended Data shows a series of measurements with increasing fluence.  Note that the rapid oscillations during the overlap of the pump and probe pulses are not indicative of dynamics in the magnetisation but arise from artefacts from the measurement that are due to the need to replicate the pump conditions of the x-ray experiment\cite{Lebedev2005,Radu2009}.  For our analysis we use only the magnitude of the pump-probe effect just after the overlap region, yielding the calibration between fluence and demagnetisation shown in Fig.~\ref{fig:fluencedependence}, which was used in the main text for the comparison between demagnetisation and mechanical angular momentum. The fluence dependence measurement confirms that the pump laser conditions of the x-ray experiment were still well below the damage threshold (occuring at about 12.5~mJ/cm$^2$ and 15\% demagnetisation) and in the linear regime of demagnetisation vs. fluence.

\section*{Longitudinal sound speed measurements}

To measure the longitudinal speed of strain waves in our film we performed pump-probe reflectivity measurements similar to Thomsen~\emph{et al.}\cite{Thomsen1986}. The pump laser had a wavelength of 800~nm, 40~fs pulse duration and a repetition rate of 250~kHz. For the probe, a second harmonic 400~nm beam was generated using a beta barium borate (BBO) crystal. We measure a time of $7.2\pm 0.1$ ps for the surface-generated acoustic wave to propagate to the substrate-Fe interface and back.  We thus estimate a longitudinal sound velocity of $4730\pm150$~m/s for the iron, where we assume that the wave propagates twice through 15.5~nm of Fe and 2.4~nm of capping material  (in agreement with the 17.9 nm stack thickness measured via TEM). Here we assumed that the capping layers consist of half epitaxial MgO with a longitudinal sound velocity along [001] of 9088~m/s~\cite{Marklund1971} and half Al with an isotropic longitudinal sound velocity of 6422~m/s~\cite{Villars2016}.

\section*{X-ray pump-probe experiment}

For the time-resolved experiment at the x-ray pump-probe endstation\cite{Chollet2015} at LCLS, an external magnetic field was used to set the magnetisation direction of the film between x-ray shots. This was achieved using a small laminated-core electromagnet driven by an IGBT-based pulser\cite{Fognini2012}. For each shot of the x-ray laser, the sample magnetisation direction was set according to a pseudorandom sequence and tagged for the data acquisition system. This prevents possible periodic disturbances like the mains frequency leaking into our magnetic difference signal. A nitrogen cryoblower was used during the measurement and the sample was cooled to 200~K; the main reason for the cooling setup was the shielding of the sample from x-ray damage in air, which we had encountered during earlier tests at the SLS synchrotron.

The x-ray probe measurement of the crystal truncation rod of the (2~2~0) in-plane Bragg peak of the film was conducted at the XPP endstation of LCLS in a grazing incidence and exit geometry. The sample was mounted vertically; the incoming angle of the x-rays was $1^\circ$ and the x-ray photon energy was 6.9~keV with an x-ray pulse rate of 120~Hz. The total integration time for the measurement was 10~hours, during which 158 pump-probe scans were completed (21 scan steps from -1~ps to 5~ps, 600 shots per step). For each x-ray shot we recorded the incident intensity $I_0$ using a diode which measures backscattering from a window. The average pulse energy from the machine was 1.8~mJ, leading to roughly $10^{9}$ photons per shot at the sample after monochromator and beamline losses.  The spectral-encoding based timing tool of the XPP endstation was used to rebin the data with 100~fs wide bins\cite{Bionta2014}. The measured average diffraction data in each bin was then normalised to the average $I_0$ for the corresponding shots. The outgoing beam was imaged using the CSPAD-140k area detector\cite{Herrmann2013}, yielding single-shot measurements of the (2~2~L) crystal truncation rod with an L-range from 0.03 to about 0.1. This range of momentum transfer is restricted on the lower end by the incidence angle of $1^\circ$ in combination with the absorption inside the sample. On the upper end, a hard limit was given by the detector area, but the strong decrease in signal-to-noise ratio with increasing L limits the useful range to the given 0.1 figure. The effective $q_z$-resolution is decreased by sample inhomogeneity, geometric and footprint effects as well as x-ray divergence.

The optical pump was delivered by a femtosecond laser working at 800~nm with a pulse duration of 40~fs and a repetition rate of 120~Hz. We set the incidence angle to 85$^\circ$ and made the pump p-polarised. For the fluence calibration, we measured the mode shape of the optical pump beam at the sample position with a CCD camera. We then recorded the total incident power for different attenuation settings, as well as the power reflected by the sample. We used the Fresnel transfer matrix method\cite{Born1999} and the known sample structure to calculate the power absorbed and reflected by the Fe layer as a function of the angle of incidence. The optical constants for this procedure were taken from literature\cite{Palik1991,Ordal1988}; the best fit to our recorded data are an incident fluence of $(8.0\pm0.3)~\text{mJ/cm}^2$ at an angle of $(85.64\pm0.05)^\circ$, with $(2.7\pm0.1)~\text{mJ/cm}^2$ absorbed by the iron film. The uncertainties in the optical constants and thicknesses do not contribute a significant error for these calculations; instead, the uncertainties in fluence are dominated by our estimate of how well pump and probe were overlapped in the x-ray and MOKE measurements.

\section*{X-ray detection of surface strain and transverse displacements}

According to the kinematic theory of x-ray diffraction the scattering intensity $I_s$ for a given momentum transfer $\mathbf{Q}$ from a crystal with a monoatomic basis is
\begin{equation}
\frac{I_s}{I_0} \propto \left|\sum_\mathbf{R} f_0 e^{i \mathbf{r}(\mathbf{R}) \cdot \mathbf{Q}}\right|^2
\end{equation}
where the sum runs over all equilibrium positions $\mathbf{R}$ of the atoms, and $\mathbf{r}(\mathbf{R})$ is an instantaneous position of the atom near $\mathbf{R}$.  Here $f_0$ is the atomic scattering factor for the basis atom.  If we now write $\mathbf{r} = \mathbf{R} + \mathbf{u}$ and $\mathbf{Q} = \mathbf{G} + \mathbf{q}$, where $\mathbf{G}$ is a reciprocal lattice vector close to $\mathbf{Q}$, we have
\begin{equation}
\frac{I_s}{I_0} \propto |F_0|^2 - 2\;\operatorname{Im}\left[ F_0^* f_0 \sum_R (\mathbf{u}\cdot \mathbf{G}) e^{i \mathbf{R} \cdot \mathbf{q}}\right] - 2\;\operatorname{Im}\left[ F_0^* f_0 \sum_R (\mathbf{u}\cdot \mathbf{q}) e^{i \mathbf{R} \cdot \mathbf{q}}\right] \label{eq:xraydiff} 
\end{equation}
where we have expended the complex exponential and kept terms up to first order in $\mathbf{u}$; $\operatorname{Im[\;]}$ takes the imaginary part of a complex expression.  Here $F_0 = \sum_R f_0 e^{i \mathbf{R \cdot \mathbf{q}}}$ is the structure factor in the absence of strain.  For our experiment $\mathbf{G}$ is directed within the surface plane and $\mathbf{k}$ is directed along the outward surface normal.  Let us define the $z$ axis as the surface normal, and write $u_T = \mathbf{u}\cdot \mathbf{G}$ as the component of the displacement along $\mathbf{G}$.
The second term of Eq.~\ref{eq:xraydiff} is  proportional to the imaginary part of the Fourier transform of $u_T$, and the third term is proportional to the imaginary part of the Fourier transform of  $u_z q_z$.    For coherent modes these terms contribute time-oscillating contributions to the x-ray diffraction signal with an amplitude proportional to the appropriate k-resolved component of the displacement field. The weighting factors are, however, quite different:  the signal from the $u_T$ component is enhanced by a factor of $G/q_z$, which for our experiment ranges from 28 to 94 as we were measuring in the (2 2 L) CTR with a useable L-range of 0.03--0.1.

\section*{Connection of transverse strain to angular momentum}

In the main manuscript we show that an antisymmetric contribution to the stress tensor that is proportional to the time derivative of the magnetisation leads to a transverse strain wave from the surface.  Our experimental data show the presence of such a wave.  Here we argue that our observations cannot be explained by any other known physical effect.

One of the basic theorems of continuum mechanics is that the conservation of mechanical angular momentum implies that the stress tensor in a material is symmetric, in other words $\sigma_{ij} = \sigma_{ji}$ \cite{Lai2009}.  Antisymmetric components to the stress violate this equality and necessarily lead to changes in the mechanical angular momentum.  Thus, we can reformulate the question of whether our transverse strain wave is the result of processes that conserve angular momentum to an equivalent question of whether any known physical mechanism that induces a symmetric stress tensor can result in our observation of transverse strain.   

We first note that in our experiment we subtract data taken at equal and opposite values of the initial magnetisation $\mathbf{M}$.  Thus, to explain our data we need a contribution to the stress tensor that depends on the magnetisation or on some time derivative of it.  The only previously known mechanism for this (besides the rotational coupling identified with the Einstein-de Haas effect) is magnetostriction, where the stress tensor depends on the instantaneous value of $\mathbf{M}$ itself.  Since ferromagnetic $\alpha$-iron has inversion symmetry, the magnetostrictive stress depends on $\mathbf{M}$ only in even orders.  Thus any magnetostriction component will be subtracted our from the signal in our data analysis.

Even if small systematic errors in the magnetisation reversal were to lead to a small magnetostrictive contribution to our signal, the stress would follow the temporal dynamics of the magnetisation $\mathbf{M}(t)$ and show a step-like behaviour over our measurement window.  The resulting dynamics when resolved in momentum space would have a cosine-like phase similar to what we see in the longitudinal strain dynamics.  In our experiment we observe clearly a sine-like phase to the oscillations, indicating that the stress inducing the dynamics lasts for a time that is short compared with the period of the Fourier components we see in our experiment.  

Finally, one could also imagine that transverse strain is generated by the substrate or the neighbouring capping layer.  These materials are, however, optically transparent, inversion symmetric, and nonmagnetic. It is therefore difficult to see how the pump could induce a strong change, and especially how the sign of the transverse strain would depend on the initial magnetisation direction in the iron film.  

\section*{Simulations and Fitting}

To properly model our x-ray diffraction experiment, the continuum model of the ultrafast Einstein--De Haas effect needed to be discretised to the atomic scale. The starting point for the lattice-dynamics part of the atomistic simulation was a Born--von-K\'arm\'an fifth-nearest-neighbour (five-shell) general-force-constant model with empirical parameters taken from Minkiewicz, Shirane, and Nathans\cite{Minkiewicz1967}. In Extended Data Table \ref{table:fcms}, we reproduce their notation for the force-constant matrices originally taken from Woods\cite{Woods1963} and described e.g. in Flocken and Hardy\cite{Flocken1969}.
For our thin-film calculation, the three-dimensional model of harmonic springs can be reduced to a chain of layers, since there are no differential forces between atoms of the same layer. In the following, we will assume the experimental situation, i.e. a $bcc$--Fe thin film sample with crystallographic surface (0~0~1) whose magnetisation lies completely in the film plane. This film orientation and geometry makes sure that the off-diagonal elements in Extended Data Table \ref{table:fcms} do not contribute to the lattice dynamics. We call the two in-plane directions $x$ and $y$ and the out-of-plane direction, which is the direction of the chain, $z$. Both the compressive (longitudinal) and the shear (transverse) strain waves will travel along the $z$-direction.

The fundamental units of this reduced model are layers of thickness $\frac{1}{2}a$, where $a = 2.860\text{ \AA}$ is the lattice constant of \emph{bcc}-Fe\cite{Sutton1955}. The independent dynamic coordinates of each layer $N$ are the displacements from the equilibrium position, $\vec{u}_N = (u_{xN},u_{yN},u_{zN})$. As stated above, no relative movement of the atoms within one layer is needed, since all forces -- both external and internal -- are identical for atoms of the same layer. The equation of motion for layer N can thus be written:
\begin{equation}
m_N\;\ddot{\vec{u}}_N = \sum\limits_{\substack{n=-3\\(N+n) \in [N_\text{min},\,N_\text{max}]}}^{3} \sum\limits_{j=x,y,z} k_{jn} \left( u_{j(N+n)} - u_{jN}\right) \hat{e}_j
\end{equation}
The number $\Delta{}N$ of neighbouring layers to be taken into account and the spring constants $k_{N\pm\Delta{}N}$ are determined from the 3D--model. For the fifth-neighbor iron model we find $\Delta{}N = 3$, and the effective spring constants for the layers are given as follows (cf. Extended Data Table \ref{table:layerks}):
\begin{align}
k_{x\pm{}3} =  k_{y\pm{}3} &= 4\cdot\beta_4\\
k_{z\pm{}3} &= 4\cdot\alpha_4\\
k_{x\pm{}2} =  k_{y\pm{}2} &= \beta_2 + 2\cdot\alpha_3 + 4\cdot\alpha_5 \\
k_{z\pm{}2} &= \alpha_2 + 4\cdot\beta_3 + 4\cdot\alpha_5\\
k_{x\pm{}1} =  k_{y\pm{}1} &= 4\cdot\alpha_1 + 4\cdot\alpha_4 +  4\cdot\beta_4\\
k_{z\pm{}1} &= 4\cdot\alpha_1 + 8\cdot\beta_4
\end{align}

For the compressive wave, we used the absorbed fluence of the iron film and the heat capacity of $bcc$-Fe compiled by Dinsdale\cite{Dinsdale1991} to calculate the thermal effect of the laser pump. The result is a final temperature of 651~K when starting at 200~K. Literature values were used for the linear thermal expansion coefficient of iron\cite{Nix1941} and the elastic constants\cite{Adams2006}; the lateral clamping of the film (when pump laser spot radius / $c_{\text{sound}} < t, \approx$ 100~ns in our case) was taken into account\cite{Lindenberg2001}.

For the transverse wave, the density\cite{Sutton1955,Meija2016} and saturation magnetisation\cite{Crangle1971} of \emph{bcc}-iron were taken from literature (for films thicker than 10~nm, bulk values are appropriate\cite{Vaz2008}), and we then discretised the spatial gradient of the initial magnetisation. As the shape of the gradient is not known precisely, there is seemingly a bit of freedom here: In one extreme, all layers of the film are initially magnetised the same, right up to the surface layer, with the magnetisation dropping to zero right at the surface over less than one unit cell. One could also imagine that the initial magnetisation reduces more smoothly over a few layers close to the surface. We modelled the demagnetisation-induced lattice dynamics for different initial cases, from a half--unit-cell sharp drop at the surface to a gradual decrease over the top 10 monolayers. While the detailed shape of the strain wave in the lattice dynamics model is slightly different, once the x-ray diffraction signal in the region of interest is calculated (i.e. in the (2~2~L) CTR with L between 0.03 and 0.1), all cases are virtually indistinguishable. This makes sense since an out-of-plane component of the momentum transfer of up to 0.1 reciprocal lattice units implies that the spatial resolution in out-of-plane direction cannot be better than about 10 monolayers.

The substrate and the capping layers of the film were treated in the model by modifying the masses of the respective parts of the linear chain such that the correct volumetric density of the respective material ($\textrm{MgAl}_2\textrm{O}_4$, MgO, Al) results. In a second step, the spring constants inside those layers were all scaled so that the correct longitudinal speed of sound for the material is reproduced. At the interfaces between dissimilar materials, the averages of the spring constants on both sides were taken. This procedure leads to the correct amount of reflection and transmission of the strain wave at the interface with the substrate, and the capping layers are fairly thin and light, such that small details of the propagation inside them do not lead to a strong effect on the dynamics in the iron film, which is what the diffraction experiment is sensitive to.

Example snapshots of the transverse components of the displacement, strain and velocity are shown in figure~\ref{fig:strainwaves} for various times, as a function of depth $z$.  The transverse velocity profile is directly related to the mechanical component of the angular momentum in the probed volume via $L_\textrm{mech} = A \int z\,\rho\,v_T(z) dz$, where $\rho$ is the material density and $A$ is the effective area of the probe.

In the last step of the simulation, the calculated real-space trajectories of all the iron layers of the simulation were mapped to the expected x-ray diffraction signal in the (2~2~L) truncation rod using a simple kinematic diffraction model with extinction.

The multi-step simulation procedure of lattice dynamics and x-ray diffraction cannot directly be fitted to the measured data. Instead, we calculated the expected magnetic difference signal for different discrete settings of demagnetisation time (0, 10, 25, 50, 100, 250, 500, 1000 and 2500~fs, and in the range 100--400~fs in 20~fs steps) and magnitude (0 to 30\% in 1\% steps). The magnetic difference signal of each simulation run, normalised per $q_z$-band to the intensity before time zero and corrected for the Debye-Waller drop, was compared to the experimental data and the goodness of fit was assessed with a $\chi^2$-measure. Only $q_z$-slices with sufficient signal-to-noise ratio were included. The best $\chi^2$ value was found at 200~fs and at a demagnetisation magnitude of 8\% of the saturation magnetisation (which corresponds to 80\% of the demagnetisation seen by MOKE, which was 10\% under the same conditions). The full $\chi^2$-maps for both time ranges are displayed in Fig.~\ref{fig:fom}.

For the simulation-independent extraction of the frequencies in Fig.~\ref{fig:acoustic}, the experimental data was binned into $q_z$ slices. The initial drop was cut off and the remaining curvature of the trace was corrected with a quadratic background subtraction. The frequencies were extracted by Fast Fourier transforms, using zero-padding and a Chebychev window with its width set to the length of the unpadded data. A Gaussian fit was finally used to extract the main frequency component of the FFT. The width of the Gaussian was constrained to less than 0.7~THz in order to correctly find the acoustic peak and not fit extremely wide peaks at very high frequencies where the transverse data are most noisy. This processing chain, optimised for the experimental data, was kept unchanged for the simulated data as verification.

In Fig.~\ref{fig:results}, the error bars for the noise-free simulated data are the asymptotic errors from the Gaussian fits to the peaks in the FFT spectra of the $q_z$ bands. We then performed a single weighted fit to a linear dispersion function $\omega = v q$ to extract the sound velocity. For the simulations the uncertainties are the asymptotic errors of this fit. For the experimental data we performed a bootstrapping procedure to estimate the uncertainties in the dispersion slopes.  For the bootstrapping, we take as a starting point the measured time traces integrated over the selected ranges of momentum transfer. We then constructed 1000 virtual new datasets by adding Gaussian distributed noise to each data point, with a $1\sigma$ width equal to the standard error of each point as determined from 158 real independent scans. We then perform the same analysis procedure as for the simulated data on each of these virtual datasets to get a distribution of results for the speed of sound. Both distributions for the speed of sound show a nearly Gaussian-shaped profile, although the difference measurements have a long but very small asymmetric tail extending to higher frequencies.  We estimate the most probable value and the uncertainties in these distributions by fitting them to a Gaussian profile and report the mean value with uncertainties corresponding to $1\sigma$ of the fitted Gaussian shape.
\end{methods}


\providecommand{\noopsort}[1]{}\providecommand{\singleletter}[1]{#1}%

\begin{addendum}
 \item[Data availability]
 Raw data were generated at the LCLS large-scale facility, and intermediate datasets were generated during analysis. The raw and intermediate datasets are available from the corresponding author on reasonable request.
 \item[Code availability]
Data processing codes and simulation codes are available from the corresponding author on reasonable request.
\end{addendum}


\newpage
\FloatBarrier

\begin{figure}
 \includegraphics[width=0.85\textwidth]{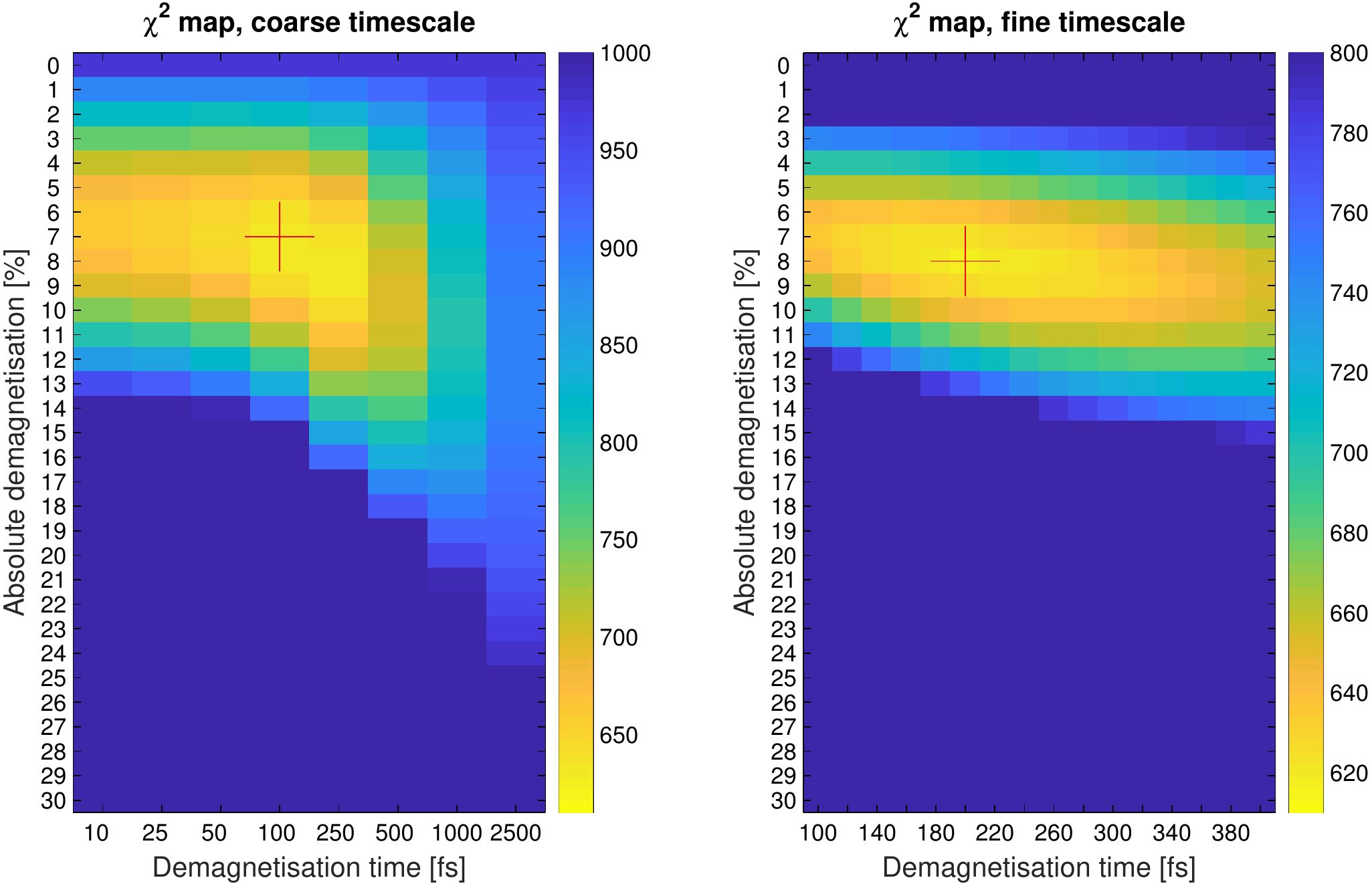}
 \caption{\textbf{Goodness of fit assessed by $\chi^2$ for different combinations of demagnetisation time and magnitude.} The left panel shows has a coarse, logarithmic axis for the demagnetisation time, the right panel has a linear scale for a more precise determination of the optimum. The optima are indicated by the red crosses; for the coarse scale, the best $\chi^2$ of 628.9 is reached for the simulation with 100~fs demagnetisation time and 7\% magnitude. For the fine scale, the best values are 200~fs and 8\%, with a $\chi^2$ of 618.8. The traces in Fig.~\ref{fig:results} were generated using these latter parameters.}
 \label{fig:fom}
\end{figure}

\begin{figure}
 \includegraphics[width=0.85\textwidth]{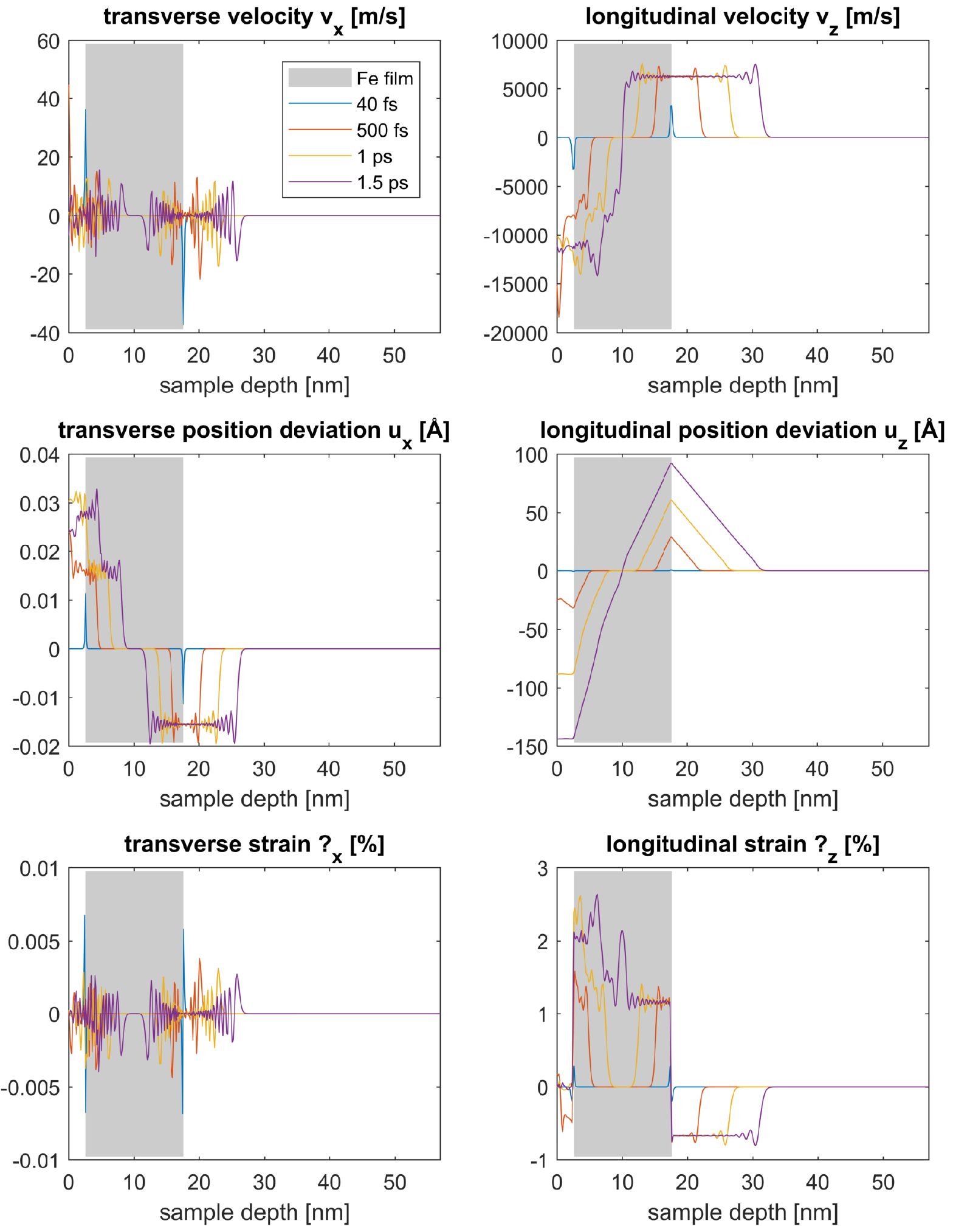}
 \caption{\textbf{Snapshots of simulated velocity, displacement, and strain in transverse and longitudinal direction at different times.} The iron film is indicated by grey shading; to the left are the Al and MgO capping layers, to the right is the MgAl$_2$O$_4$ substrate.}
 \label{fig:strainwaves}
\end{figure}

\FloatBarrier

\begin{table}
\renewcommand{\arraystretch}{1.2} 
\begin{tabular}{@{}c@{\hskip 6pt}c@{\hskip 6pt}c@{}}
\toprule
Atom position				&Force constant matrix & \emph{bcc} Fe, 300K\\
\midrule
 $\frac{a}{2}(1,1,1)$ & $ \left( \begin{array}{ccc}\alpha_1 & \beta_1  & \beta_1 \\ \beta_1  & \alpha_1 & \beta_1  \\ \beta_1  & \beta_1  & \alpha_1\end{array} \right) $
 & $ \begin{array}{@{}c@{\,}c@{\,}r@{}c@{}c@{}}\alpha_1 &=& 16&.&88\frac{\text{N}}{\text{m}} \\ \beta_1 &=& 15&.&01\frac{\text{N}}{\text{m}}\end{array} $\\
\midrule
 $\frac{a}{2}(2,0,0)$ & $ \left( \begin{array}{ccc}\alpha_2 & 0        & 0       \\ 0        & \beta_2  & 0        \\ 0        & 0        & \beta_2 \end{array} \right) $
 & $ \begin{array}{@{}c@{\,}c@{\,}r@{}c@{}c@{}}\alpha_2 &=& 14&.&63\frac{\text{N}}{\text{m}} \\ \beta_2 &=& 0&.&55\frac{\text{N}}{\text{m}}\end{array} $\\
\midrule
 $\frac{a}{2}(2,2,0)$ & $ \left( \begin{array}{ccc}\alpha_3 & \gamma_3 & 0       \\ \gamma_3 & \alpha_3 & 0        \\ 0        & 0        & \beta_3 \end{array} \right) $
 & $ \begin{array}{@{}c@{\,}c@{\,}r@{}c@{}c@{}}\alpha_3 &=& 0&.&92\frac{\text{N}}{\text{m}} \\ \beta_3 &=& -0&.&57\frac{\text{N}}{\text{m}} \\ \gamma_3 &=& 0&.&69\frac{\text{N}}{\text{m}}\end{array} $\\
\midrule
 $\frac{a}{2}(3,1,1)$ & $ \left( \begin{array}{ccc}\alpha_4 & \delta_4 & \delta_4\\ \delta_4 & \beta_4  & \gamma_4 \\ \delta_4 & \gamma_4 & \beta_4 \end{array} \right) $
 & $ \begin{array}{@{}c@{\,}c@{\,}r@{}c@{}l@{}}\alpha_4 &=& -0&.&12\frac{\text{N}}{\text{m}} \\ \beta_4 &=& 0&.&03\frac{\text{N}}{\text{m}} \\ \gamma_4 &=& 0&.&52\frac{\text{N}}{\text{m}}\\ \delta_4 &=& 0&.&007\frac{\text{N}}{\text{m}}\end{array} $\\
\midrule
 $\frac{a}{2}(2,2,2)$ & $ \left( \begin{array}{ccc}\alpha_5 & \beta_5  & \beta_5 \\ \beta_5  & \alpha_5 & \beta_5  \\ \beta_5  & \beta_5  & \alpha_5\end{array} \right) $
 & $ \begin{array}{@{}c@{\,}c@{\,}r@{}c@{}c@{}}\alpha_5 &=& -0&.&29\frac{\text{N}}{\text{m}} \\ \beta_5 &=& 0&.&32\frac{\text{N}}{\text{m}}\end{array} $\\
\bottomrule
\end{tabular}
\caption{\label{table:fcms}%
\textbf{Force constant matrices for $bcc$ metals in the 5-shell Born--von K\'{a}rm\'{a}n model.} The values in the right column are from Minkiewicz \emph{et al.}\cite{Minkiewicz1967}. The notation is reproduced from 
Flocken and Hardy\cite{Flocken1969}.}
\end{table}

\begin{table}
\renewcommand{\arraystretch}{1.2} 
\begin{tabular}{@{}ccccccc@{}}
\toprule
\shortstack[c]{Neighbour \\ class}          & \shortstack[c]{Specific \\ neighbours} & \shortstack[c]{In \\ layer} & \shortstack[c]{\# of \\ atoms} & $k_x$        & $k_y$        & $k_z$        \\
\midrule
\multirow{2}{*}{(1,1,1)} & \phantom{m}$(\pm1,\pm1,\phantom{\pm}1)$\phantom{m}&$N+1$& 4       & $\alpha_1$ & $\alpha_1$ & $\alpha_1$ \\
                         & $(\pm1,\pm1,-1)       $&$N-1$& 4       & $\alpha_1$ & $\alpha_1$ & $\alpha_1$ \\
\midrule
\multirow{4}{*}{(2,0,0)} & $(\phantom{\pm}0,\phantom{\pm}0,\phantom{\pm}2)           $&$N+2$& 1       & $\beta_2$  & $\beta_2$  & $\alpha_2$ \\
                         & $(\pm2,\phantom{\pm}0,\phantom{\pm}0)          $&$N$& 2       & $\alpha_2$ & $\beta_2$  & $\beta_2$  \\
                         & $(\phantom{\pm}0,\pm2,\phantom{\pm}0)          $&$N$& 2       & $\beta_2$  & $\alpha_2$ & $\beta_2$  \\
                         & $(\phantom{\pm}0,\phantom{\pm}0,-2)           $&$N-2$& 1       & $\beta_2$  & $\beta_2$  & $\alpha_2$ \\
\midrule
\multirow{5}{*}{(2,2,0)} & $(\pm2,\phantom{\pm}0,\phantom{\pm}2)          $&$N+2$& 2       & $\alpha_3$ & $\beta_3$  & $\alpha_3$ \\
                         & $(\phantom{\pm}0,\pm2,\phantom{\pm}2)          $&$N+2$& 2       & $\beta_3$  & $\alpha_3$ & $\alpha_3$ \\
                         & $(\pm2,\pm2,\phantom{\pm}0)        $&$N$& 4       & $\alpha_3$ & $\alpha_3$ & $\beta_3$  \\
                         & $(\pm2,\phantom{\pm}0,-2)         $&$N-2$& 2       & $\alpha_3$ & $\beta_3$  & $\alpha_3$ \\
                         & $(\phantom{\pm}0,\pm2,-2)         $&$N-2$& 2       & $\beta_3$  & $\alpha_3$ & $\alpha_3$ \\
\midrule
\multirow{6}{*}{(3,1,1)} & $(\pm1,\pm1,\phantom{\pm}3)        $&$N+3$& 4       & $\beta_4$  & $\beta_4$  & $\alpha_4$ \\
                         & $(\pm3,\pm1,\phantom{\pm}1)        $&$N+1$& 4       & $\alpha_4$ & $\beta_4$  & $\beta_4$  \\
                         & $(\pm1,\pm3,\phantom{\pm}1)        $&$N+1$& 4       & $\beta_4$  & $\alpha_4$ & $\beta_4$  \\
                         & $(\pm3,\pm1,-1)       $&$N-1$& 4       & $\alpha_4$ & $\beta_4$  & $\beta_4$  \\
                         & $(\pm1,\pm3,-1)       $&$N-1$& 4       & $\beta_4$  & $\alpha_4$ & $\beta_4$  \\
                         & $(\pm1,\pm1,-3)      $&$N-3$& 4       & $\beta_4$  & $\beta_4$  & $\alpha_4$ \\
\midrule
\multirow{2}{*}{(2,2,2)} & $(\pm2,\pm2,\phantom{\pm}2)        $&$N+2$& 4       & $\alpha_5$ & $\alpha_5$ & $\alpha_5$ \\
                         & $(\pm2,\pm2,-2)       $&$N-2$& 4       & $\alpha_5$ & $\alpha_5$ & $\alpha_5$ \\
\bottomrule
\end{tabular}
\caption{\label{table:layerks}%
\textbf{Force constant contributions for a chain of layers along $z$.} The classes of neighbour atoms, and their multiplicitites and contributions to the force constants between layers are given for the principal directions.
}
\end{table}

\end{document}